\documentclass{ws-procs9x6}
\usepackage{graphicx}

\begin{document}

\title{The reaction $\bar p p\rightarrow \pi^+\pi^-$: relativistic aspects and final-state interaction in the 
       pion wave functions \footnote{this talk is a condensed version of reference [1] and as of yet not 
       published work.}}

\author{\underline{B.~El-Bennich$^{\,1}$}, W.M.~Kloet$^{\,1}$, and B.~Loiseau$^{\,2}$}
\address{$^1$~Department of Physics and Astronomy, Rutgers University, \\
         136 Frelinghuysen Road, Piscataway, New Jersey 08854 \\
         $^2$ LPNHE (Groupe Th\'eorie), Universit\'{e} P.\& M. Curie, 4 Place Jussieu, 75252 Paris Cedex 05, France}

\maketitle

\abstracts{We use a distorted wave approximation approach which includes  $^3P_0$ and $^3S_1$ quark-antiquark annihilation 
           mechanisms to reproduce the data set from LEAR on $\bar p p\rightarrow \pi^+\pi^-$ in the range from 360 to 1550 MeV/c. 
           Improvements of the model are sought by implementing final-state interactions of the pions and by observing that the 
           annihilation is too short-ranged in earlier attempts to describe the data. While the former improvement 
           is due to to the final-state $\pi\pi$ wave functions solely, the latter one originates from quark wave functions 
           for proton, antiproton, and pions with radii slightly larger than the respective measured charge radii. 
           This increase in hadron radius, as compared with typically much smaller radii used before in the quark model, 
           increases the annihilation range and thereby the amplitudes for $J\ge2$ are much higher. Finally, given the very 
           high kinetic energy of the final pions, we investigate the role of relativistic 
           corrections in the pion wave functions when boosted into the center-of-mass frame.}

\section{Introduction}

The very accurate set of data from the LEAR experiment \cite{hasan92} on $\bar pp\to\pi^+\pi^-$ measuring the 
differential cross section and analyzing power from 360 to 1550 MeV/c is still a challenge for theoretical models 
after more than a decade. Large variations are observed in the analyzing power $A_{0n}$ as a function of angle at all 
energies, indicating presence of several partial waves already at low energies. 
However, recent model calculations \cite{moussallam83,mull91,mull92,bathas93,muhm96,yan96} lead to scattering amplitudes 
which are strongly dominated by total angular momentum $J=0$ and $J=1$. The reason for this is the choice of a rather 
short range annihilation mechanism. The short range of the annihilation in the model calculations originates from 
the dynamics of baryon exchange in Refs.~\cite{moussallam83,mull91,mull92,yan96} or from required overlap of 
quark and antiquark wave functions for proton and antiproton in Refs.~\cite{bathas93,muhm96}. 
On the other hand the experimental data on differential cross sections as well as those on asymmetries point to 
a significant $J=2$, $J=3$ and even higher $J$ contributions~\cite{oakden94,hasan94,kloet96,martin97}. 
 All above mentioned models, for this reaction, use a distorted wave approximation (DWA). In order to calculate the 
$\bar p p\rightarrow \pi^+\pi^-$ amplitudes we use the  initial coupled spin-triplet $\Psi_{\bar pp}({\mathbf r})$
wave functions in configuration space as provided by the Paris 1998 $\bar NN$ potential \cite{elbennich98}. The transition 
operator $O({\mathbf r}',{\mathbf r})$ is computed from quark and antiquark diagrams in which a $\bar qq$ pair is annihilated
into either an {\em effective} vacuum $^3P_0$ or ''gluon'' like $^3S_1$ state. In both cases momentum is transferred
from the annihilation vertex to the spectator quarks. The final ingredient is the $\pi\pi$ wave function 
$\Psi_{\pi\pi}({\mathbf r}')$  parameterized in terms of $\pi\pi$ phase shifts and inelasticities. The complete 
scattering amplitude $T$ using the DWA is then
\begin{equation}
	T =\int d{\mathbf r}'d{\mathbf  r}\ \Phi_{\pi\pi}({\mathbf r}')\,O({\mathbf r}',{\mathbf r})
	\Psi_{\bar pp}(\mathbf r).
	\label{1}
\end{equation}

\section{Observables and $\pi\pi$ final-state interaction}
The reaction $\bar pp\to \pi^+\pi^-$ can be fully described in the helicity formalism by two independent helicity amplitudes 
$F_{++}(\theta)$ and $F_{+-}(\theta)$. The angle $\theta$ is the c.m. angle between the outgoing $\pi^-$ and the incoming $\bar p$.
The two observables measured at LEAR are \cite{hasan92}
\begin{equation}
\label{3}
\frac{d\sigma}{d\Omega}=\frac{1}{2}(\vert F_{++}\vert^2+\vert F_{+-}\vert^2), \qquad
      A_{0n} \frac{d\sigma}{d\Omega}=\mathrm{Im}\,(F_{++}F_{+-}^*),
\end{equation}
where the helicity amplitudes are obtained from the DWA method in Eq.~(\ref{1}). The final $\pi\pi$ scattering wave functions 
$\Phi_{\pi\pi}(\mathbf{r'})$ supersede previously used plane waves. The elastic $\pi\pi\to \pi\pi$ amplitude is known from threshold 
up to the total relativistic $\pi\pi$ energy $\sqrt{s}= 1800$ MeV mainly from analysis of the $\pi N\to \pi\pi N$ reaction. 
The extracted $\pi\pi\to \pi\pi$ amplitudes can be parameterized in terms of phase shifts $\delta_J$ and inelasticities $\eta_J$ 
where $J=0,1,2$, and 3. This final $\pi\pi$ interaction proves to be a sensitive ingredient in the fit to the observables. 

\section{Quark wave functions and charge distribution radii}

Within the quark model the spin-momentum structure of the annihilation amplitudes is dictated by the topology of the flavor 
flux as well as by the vertices, whereas the range is determined by the overlap of quark and antiquark wave functions
for proton, antiproton and pions. These wave functions are 
\begin{equation}
\label{9}
 \psi_p({\mathbf r}_1,{\mathbf r}_2, {\mathbf r}_3)= N_p\, \exp \left[ -\frac{\alpha}{2}
 \sum_{i=1}^3({\mathbf r}_i-{\mathbf r}_p)^2 \right]\times\!\chi_{_p}\mathrm{(spin, isospin, color)},
\end{equation}
where  ${\mathbf r}_i$ are the quark coordinates and ${\mathbf r}_{p}$ the proton coordinate. For the antiproton
the antiquarks are ${\mathbf r}_4, {\mathbf r}_5$ and ${\mathbf r}_6$.
An $S$-wave meson intrinsic wave function is:
\begin{equation}
\label{10}
 \phi_{\pi}({\mathbf r}_1, {\mathbf r}_4)=N_{\pi}\, \exp \left[ -\frac{\beta}{2}
 \sum_{i=1,4}( {\mathbf r}_i-{\mathbf r}_{\pi})^2 \right]\times\!\chi_{_{\pi}} \mathrm{(spin, isospin, color).}
\end{equation}
Here ${\mathbf r}_1$ and $ {\mathbf r}_4$ are the quark and antiquark coordinates of one pion, respectively. The coordinate of the 
pion is ${\mathbf r}_{\pi}$. In a fit to a representative set of $d\sigma/d\Omega$ and $A_{0n}$ data \cite{hasan92} at five 
energies the size parameters $\alpha$ and $\beta$ take on values which correspond to $\langle r^2_{p}\rangle^{1/2}=0.91$~fm and 
$\langle r^2_{\pi}\rangle^{1/2}=0.71$~fm, respectively. This is within 7\% of the measured charge distribution radii values found in 
the literature \cite{pdg02,amendolia} and considerably larger than the proton, antiproton and pion radii used before.

\section{Relativistic corrections}

Final states with kinetic energies much higher that the pion rest mass are produced in the LEAR
experiment. For an incoming antiproton with $p_{\mbox{\scriptsize lab}}=800$ MeV/c one obtains a relativistic factor 
$\gamma=E_{\mathrm{cm}}/2m_{\pi}c^2=7.2$ for the outgoing pions. The Gaussian spheres of Eqs.~(\ref{9},\ref{10}) are appropriate in 
the rest frame of the pions. However, the transition amplitudes and therefore the observables (\ref{3}) are calculated in the 
c.m. frame. The $\pi\pi$ wave functions must be Lorentz transformed from their rest frame into the relevant c.m. frame. The boosts 
``flatten'' the pion wave functions. After a Lorentz boost along the relative pion coordinate  ${\bf r}_{\pi}$ the exponential part 
of the wave functions becomes
\begin{equation}
 \exp \left [ -\frac{\beta}{2}\!\sum_{i=1,4}\!\left (
 (\mathbf{r}_{i} - {\bf r}_{\!\pi} )_{\perp}^2\!+ \gamma^2 (\mathbf{r}_{i} - {\bf r}_{\!\pi})_{\|}^2 \right ) \right ]~. 
\end{equation} 
These modified wave functions can be shown to considerably change the angular distribution due to
additional terms in the annihilation amplitude. Also, as for the changes in the last two sections, partial waves for $J\geq 1$ are 
strongly enhanced.

\section{Results}

Some partial results for $d\sigma/d\Omega$ and $A_{0n}$ at $T_{\mbox{\scriptsize lab}}=123.5$ and 219.9 MeV
($p_{\mbox{\scriptsize lab}}=497$ and 679 MeV/c) are shown in the figures above. The dotted curves correspond 
to the quark model with final $\pi\pi$ plane waves. Switching on the $\pi\pi$ final state interactions yields 
the dashed and furthermore increasing the hadronic radii the solid curves. Relativistic effects have not yet been 
included. 
\begin{figure}[t]
\begin{center}
\includegraphics[angle=90,scale=0.5]{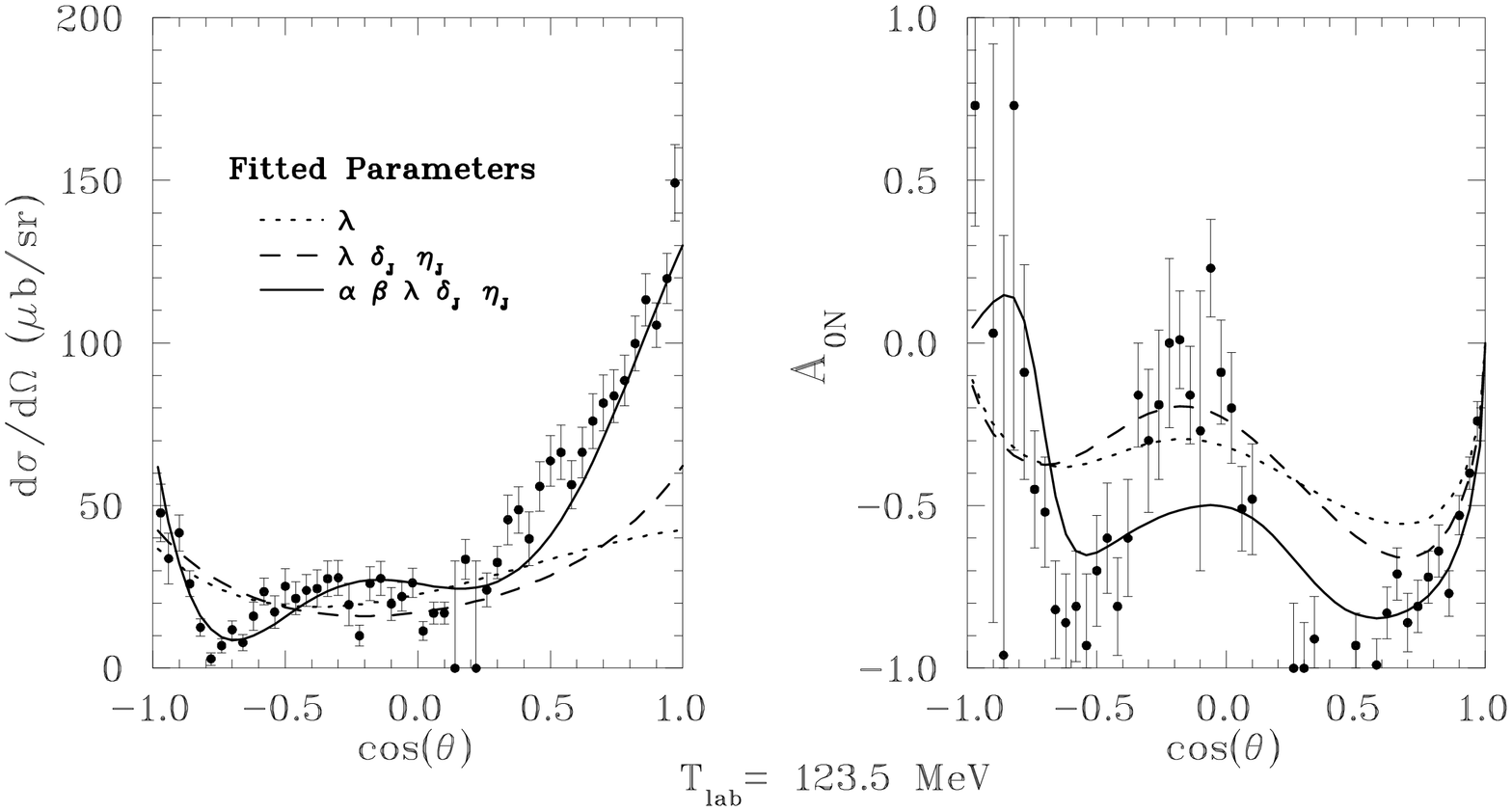}
\includegraphics[angle=90,scale=0.5]{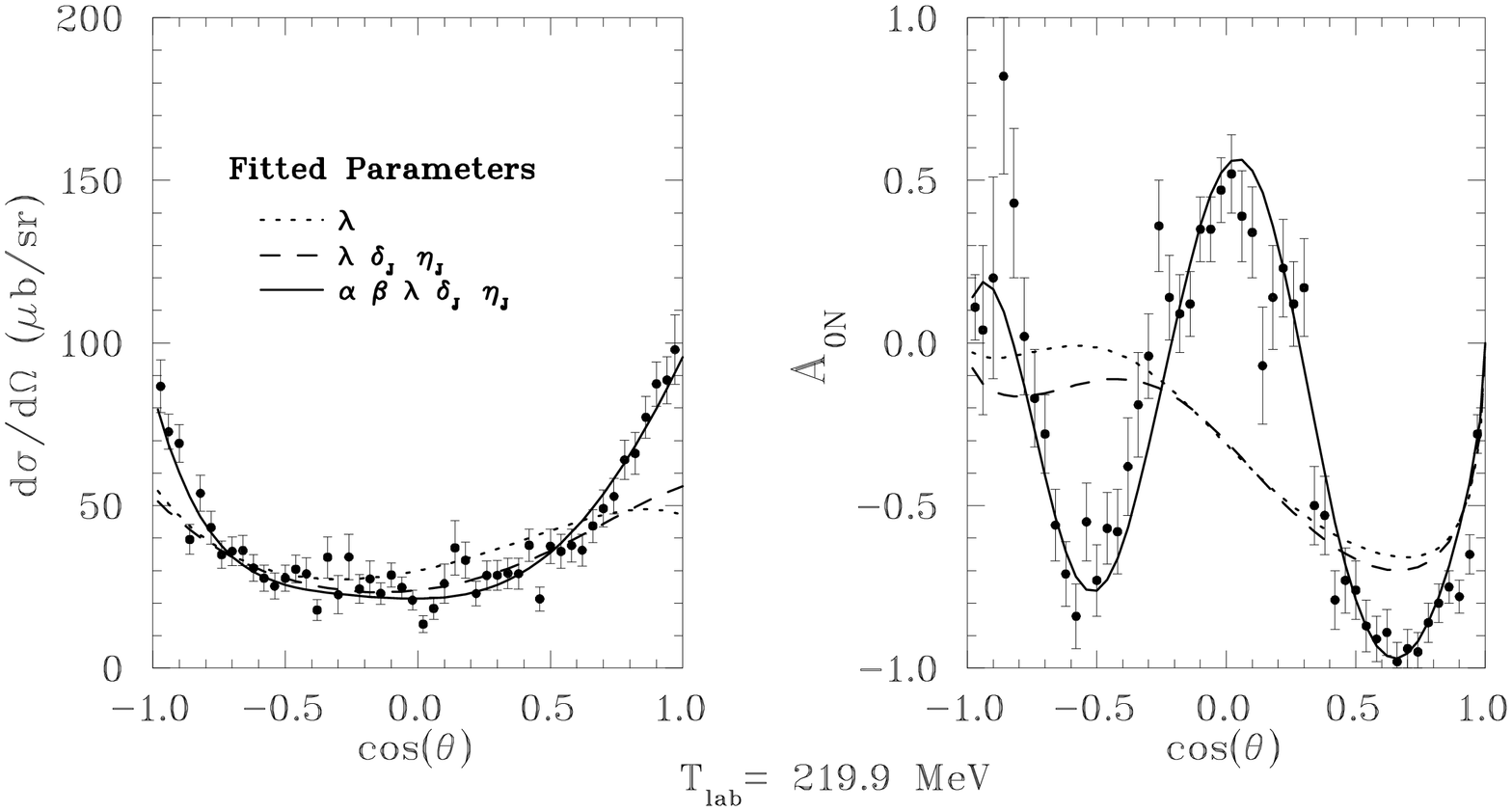}
\end{center}
\end{figure}

\section*{Acknowledgment}
B.E. would like to thank the organizers of the $N^*$2002 conference in Pittsburgh for the financial support 
and the opportunity to give this talk.

\end{document}